\documentstyle[aps,prb,epsf,floats,twocolumn]{revtex}

\begin{document}

 \bibliographystyle{prsty}


 \title{
  Resonant tunnelling and quenching of tunnel splitting in Wess-Zumino
nanospin systems}

 \author{Soo-Young Lee$^a$ and Sahng-Kyoon Yoo$^b$}

 \address{$^a$ Department of Physics, Kyungnam University, Masan 631-701, Korea}
 \address{$^b$ Department of Physics, Chongju University, Chongju 360-764, Korea\\
\smallskip
{\rm (\today)}
\bigskip\\
\parbox{14.2 cm}
{\rm
We investigate the energy spectrum of the biaxial spin systems with magnetic field
along the hard anisotropy axis by using the complex periodic orbit theory.
All important features of the system appearing in whole energy range, such as
oscillations of level splittings due to Wess-Zumino effect and their absence
at higher magnetic field etc., can be completely understood within this semiclassical
scheme. We find out that the fields at which the tunnelling quenches do not shift
at higher energy levels and the absences of the quenching at higher magnetic field have
their origin in an exact coincidence of the quenching field with the field of resonant tunnelling.
 Based on the result, we propose that the complete cancellation of quenching with resonant
tunnelling would be a general property of  Wess-Zumino tunnelling systems.
\smallskip
 \begin{flushleft}
 \end{flushleft}
}}

\maketitle

\newpage

Recently, the oscillations of the level splitting of the nanospin systems with magnetic
field along the hard axis  has been studied by many authors theoretically\cite{garg93,garg99,kou99,yoo99} and 
experimentally\cite{wern99}. This oscillation phenomenon is semiclassically understood as a result of
the interference between instanton and anti-instanton  in the Euclidean space\cite{garg93}.
However, although this instanton method gives exact evaluation of the oscillation
of the ground level splitting, it does not give a satisfactory explanation of  
how  the quenching disappears at higher magnetic field and how higher level
splitting behaves with magnetic field.

On the other hand, there have been many works about complex periodic orbit theory
for the tunnelling systems\cite{rob89,rob90,mil79,brin83}. 
This theory is based on the extension of well-known
periodic orbit theory\cite{gutz,berry} by including tunneling paths going through energy barrier.
The tunnelling path have a minimal Euclidean action with a given energy,
so they can be regarded as a 'bounce'\cite{cole77}, 'instanton'\cite{cole79}, 
or 'periodic instanton'\cite{khle91}.
The advantage of complex periodic orbit theory is that for tunnelling systems
 it gives energy levels and their splittings throughout whole energy range.
 It is worth while to see how the geometrical phase can be incorperated with
 the complex periodic orbit theory, because, to our knowledge, for the systems
 with geometrical phase the theory has not been applied so far.

The purpose of this Letter is to apply the complex periodic orbit theory
to the nanospin system showing geometrical phase effect, 
and to give semiclassical
explanation for the existence of the level-splitting oscillations at low magnetic field 
and the absence of the oscillations at higher magnetic field for
excited states as well as ground state.
As will be shown clearly, we find out that the mechanism of the disappearance
of quenching is the exact cancellation of quenching with the divergence
of resonant tunnelling, which is very novel feature not reported so far.
This strong correlation between quenching and resonant tunnelling
seems to be inherent  property  in Wess-Zumino tunnelling systems.

 Let us start with the Hamiltonian of system with magnetic field along the hard axis,
\begin{equation}
 {\cal H}= K_1 S_z^2 + K_2 S_y^2 - g \mu_B H S_z,
 \label{ham}
 \end{equation}
where $K_1$ and $K_2$ are anisotropy coefficients with $K_1 > K_2$, and $g$ and $\mu_B$
are the gyromagnetic ratio and the Bohr magneton, respectively.
The effctive Euclidean action for this Hamiltonian is obtained by expressing
it in $(\theta, \phi )$- representation and integrating over $\theta$\cite{chud99},
 \begin{equation}
 S_E = iS_{W}
 +S\sqrt{\lambda}\int d\tau \left [ \frac{1}{2} m(\phi) \dot{\phi}^2 +V(\phi)\right ] 
 \end{equation}
  where $S_{W}$ denotes the Wess-Zumino term which is expressed as     
\begin{equation}
S_{W} = S\int d\tau \dot{\phi} ( 1 - \frac{h}{1-\lambda \sin ^2 \phi }),
\end{equation}
and the position dependent mass and effective potential are
\begin{eqnarray}
m(\phi) &=& \frac{1}{1-\lambda \sin ^2 \phi },  \\
V(\phi) &=& \frac{1}{2} \sin ^2 \phi (1 - \frac{h^2}{1-\lambda \sin ^2 \phi }).
\label{pot}
\end{eqnarray}
Here we put   $\lambda \equiv \frac{K_2}{K_1}$ and $h=\frac{H}{H_c}$ with 
$H_c \equiv \frac{2K_1 S}{g \mu_B}$.
Note that the geometrical phase factor $S_W$ does not change under
analytic continuation of Euclidean time into real time, i.e., $\tau \rightarrow i t$.
This means that one should consider the phase effect of $S_W$ for the 
path in real space in the same way as for the tunnelling path in the Euclidean space.
This point of view is somewhat different from that of Ref.\cite{kim99} where only the phase
effect of tunnelling path is taken. The lack of the contribution of real path to the
phase effect has created unreasonable oscillation of level splitting with temperature.

Now, we apply complex periodic orbit theory to this effective Hamiltonian system
with Wess-Zumino term.
In the regime of $h<h_1 \equiv 1-\lambda$, as depicted in Fig. 1 (a),
 the potential shows a simple barrier between $\phi=0$ and $\pi$.
 Let us first consider the energy range of $E < E_b$, $E_b$ being the energy of the barrier top.
A periodic orbit starting from $\phi=\phi_1$ can be considered 
as a combination of four elements;
(tunnelling segment + tunnelling segment ), (real segment + real segment),
(tunnelling segment + real segment), and (real segment + tunnelling segment).
With a prescription for Maslov index for tunnelling segment in Ref.\cite{rob89},
we can write semiclassical trace of Green's function in terms of a periodic orbit sum as
\begin{eqnarray}
g(E) &=&\frac{T}{i}\sum_{m=0}^{1}\sum_{N=0}^{\infty} [ (e^{-\theta+i\frac{\pi}{2}})^2 +
(e^{iW-i\frac{\pi}{2}})^2  \nonumber \\
&+&(e^{-\theta+i\frac{\pi}{2}})(e^{iW-i\frac{\pi}{2}})e^{i\alpha}e^{im \pi} \nonumber \\
&+&(e^{iW-i\frac{\pi}{2}})(e^{-\theta+i\frac{\pi}{2}})e^{-i\alpha}e^{im \pi}] ^N
\end{eqnarray}
where $T$ is the period of the real orbit in the well, 
$W$ is the action for the real segment, i.e.,
$W(E) = \int_{-\phi_1}^{\phi_1} d\phi m(\phi) \frac{d\phi}{d t}$,
$\theta$ is the action for the tunnelling segment, i.e.,
$\theta(E) = \int_{\phi_1}^{\pi -\phi_1} d\phi m(\phi) \frac{d\phi}{d \tau}$,
and $\alpha = S \pi (1-\frac{h}{\sqrt{1-\lambda}})$ is the geometrical phase
for an element corresponding to $\pi$ increment in $\phi$ axis.
The first two terms correspond to the elements (tunnelling segment + tunnelling segment)
and (real segment + real segment), respectively. These elements represent
oscillatory motions confined in the wells and barriers, which do not enclose $z$-axis
and thus have no geometrical phase factor.
While, the next two terms, (tunelling segment + real segment) and (real segment +
tunnelling segment) represent  half rotations in the direction of increasing $\phi$
and decreasing $\phi$, respectively, so that they have the geometrical phase
factor $e^{i\alpha}$ or $e^{-i\alpha}$ according to the rotation direction. 
The contributions of nonperiodic orbits ending at $\phi_1 +\pi$ in the combinations
are removed by introducing $m$ sum.  
Using the same approximation as Ref.\cite{rob89}, we can write the sum as
\begin{equation}
g(E) \simeq \sum_{k} \frac{1}{E- (E_k \pm \frac{2}{T}\cos \alpha  e^{-\theta})}
\label{lessg}
\end{equation}
where $E_k$ is the EBK(Einstein-Brillouin-Keller) energies 
determined by $2W(E_k) = 2 \pi (k + 1/2)$, $k$ being
integer, and $T = \frac{d(2W)}{dE}$.  
Since the poles of $g(E)$ indicate the eigenvalues of the system, we can see
the oscillation of the energy splitting and the quenching fields are
determined by $\alpha (h) = \pi ( n + 1/2 )$, $n$ being integer. 
It is noted that these quenching fields do not
shift  in excited energy level splittings.

Now, consider the regime of $h>h_1$. In this case the barrier top becomes local
mininum and then equivalent two barriers and new meta-stable well appear as shown in Fig. 1 (b).
Eq.(\ref{lessg}) is still valid in the energy range of $E < E_m$.
Consider the energy range of $E_m < E < E_b$, $E_m$ being the local minimum energy.
In this structure one can expect that the resonant tunnelling may happen due to
the states in the meta-stable well. As will be shown below, however, the
resonant tunnelling does not take place, since it compensates the quenching of the level splittings.
As a result, there is no more oscillations of level splittings.
The periodic orbit sum can be done simply if one attaches a small correction 
\begin{equation}
\left ( 1-\frac{e^{-2\theta_m+i\pi}}{1-e^{2iW_m-i\pi}} \right ) ^{-1}
\label{small}
\end{equation}
to each segment and replace the contribution of tunnelling segment by
\begin{equation}
\frac{e^{-2\theta_m+i\pi} e^{iW_m-i\pi/2}}{1-e^{2iW_m-i\pi}}
\label{tunnel} 
\end{equation}
where $W_m (E) = \int_{\phi_2}^{\pi -\phi_2} d\phi m(\phi) \frac{d\phi}{d t}$
and $\theta_m (E) =\int_{\phi_1}^{\phi_2} d\phi m(\phi) \frac{d\phi}{d \tau}$.
The small correction, Eq.(\ref{small}), comes from the consideration of
possible paths, leaving from $\phi_1$ or $\phi_1 + \pi$ and returning back after
some oscillatory motion in the meta-stable well, and their repetition between two segments.
In fact, this small correction does not play any important role under the approximation
used in getting the final $g(E)$.
The contribution of  the new tunnelling segment, Eq.(\ref{tunnel}),
  shows all possible oscillatory motions
in the meta-stable well between two equivalent tunnelling subsegments. 
Then, the trace of Green's function can be written approximately as
\begin{equation}
g(E) \simeq \sum_{k} \frac{1}{E- (E_k \pm \frac{1}{T}
\frac{\cos \alpha}{\cos W_m}  e^{-2\theta_m})}
\label{g2}
\end{equation}
Here, it is noted that the geometrical phase effect  $\cos \alpha$
is modulated by the action of meta-stable state. 
This effect is the main point in this Letter and will be discussed in detail soon.
 Note that in the limit $E \rightarrow E_m$, Eq.(\ref{g2}) 
gives smaller energy splittings than those given by Eq.(\ref{lessg}), i.e., 
shows a discrepancy of factor 2. This fact leads to a discontinuity at $E=E_m$
in energy splittings (see Fig.3). To remove this discrepancy at $E=E_m$, one
should use an uniform approximation\cite{rob90,ber72} which may also give
a smooth behavior near $E_b$ as well as near $E_m$.
However, we do not use the uniform approximation because the simple
complex periodic orbit theory used here is sufficient for our
purpose of this Letter.

If the energy is larger than $E_b$, the particle motion is simple rotation and
the poles of the trace of Green's function appear when
\begin{equation}
S_{rot}(E) \pm 2 \alpha = 2 \pi k , \,\,\, k = integer,
\label{rot}
\end{equation}
where the action $S_{rot} (E) = \int _{0}^{2\pi} d\phi m(\phi) \frac{d\phi}{d t}$ and
this condition gives EBK energies $E_k$ for $E>E_b$.
In this energy range the energy splitting  does not come from tunnelling, but 
rather from the Zeeman effect. Indeed, this simple semicalssical evalution
of EBK energies gives exact eigenvalues for uniaxial spin system of
$ {\cal H}= S_z^2 - H S_z$ which is not  a tunnelling system but  
an example of simple rotational orbit. In this uniaxial spin system
the mass is $1/2$, the potential is $-H^2/4$, and $\alpha = \pi (S-H/2)$,
and then from Eq.(\ref{rot}) the EBK energies is given as $E_k = k^2 - k H$
where $k$ is integer or half integer according to the spinvalue $S$.

All actions, $W(E)$, $\theta(E)$, $W_m(E)$, $\theta_m(E)$, and $S_{rot} (E)$ are
analytically calculated by the complete elliptic integrals of the first and third kind\cite{byrd71}
(see Table 1).

In Fig.2 (a) the whole energy spectrum with $h$ are shown for $S=10$ case.
The solid line and the dashed line denote the barrier energy $E_b$ and the local
minimum energy $E_m$, respectively, with $h$. In order to avoid complicated view, we draw
only EBK energies for $E<E_b$ without showing the splitting of the EBK energies.
For the comparison we draw the result of direct diagonalization for the same system.
In this direct numerical calculation we take the energy such that $E=0$ at
the well minimum. In fact, the effective potential in Eq.(\ref{pot}) has been derived
after the same energy translation.
We can see that the entire behaviour of energy spectrum with $h$ is very similar
to the result of direct diagonalization except for the region of $E \simeq E_b$.
As mentioned before, an uniform approximation would give a consistent explanation
for this transition region.
In our  semiclassical evaluation of the energy spectrum the physics in the energy
range of $E > E_b$ is very different from that in the range of $E < E_b$;
the former is the Zeeman splitting and the latter is tunnelling splitting.
However, in quantum mechanical result these seem to be mixed around $E\simeq E_b$
and no abrupt change at $E = E_b$ is shown.

Fig.3 shows the lowest three level splittings, $\Delta E_0$, $\Delta E_1$, and $\Delta E_2$.
Here, the solid lines describe the results of complex periodic orbit
 theory($\Delta E_i^s$), while
the dashed lines show the result of direct diagonalization($\Delta E_i^q$).
It should be emphasized that in the evaluation of the actions related to
tunnelling segment, i.e., $\theta (E)$, $W_m(E)$, and $\theta_m(E)$, the
energy value used is not the EBK energies $E_k$,
 but $E^*_k$ determined from the condition $2 W(E^*_k) = 2 \pi k$.
This choice of energy is consistent with usual instanton
method where the ground state splitting is represented in terms of Euclidean action at
the well minimum energy $E_{min}$(action of instanton)
 instead of  that at the ground state energy $E_0$.
The discrepancies between the quantal results ($\Delta E_{i}^q$) and the semiclassical
results ($\Delta E_i^s$) would be resolved provided that the quantum fluctuations around
the tunnelling path are taken into account.
It is very interesting to see that the quenching of the level splitting does not exist
at higher $h$, and for $\Delta E_0$ the number of quenching is $S$ and
$\Delta E_i$ has $(S-i)$ quenchings in the range of $0 < h < 1$.
This can be understood if one knows the role of $\cos W_m$ in Eq.(\ref{g2}).
The factor $1/\cos W_m$ diverges at $2W_m = 2\pi (n+1/2)$, $n$ being integer,
showing the resonant tunnelling, which means that if $E^*_k$ coincides
with the EBK energy $E_j^{meta}$ in the meta-stable well the level splitting
becomes divergent. The fact that in the range of $E_m < E_{i} < E_b$
the splittings $\Delta E_i$ $(i=0,1,2,3)$ do not show any quenching
and any resonant tunnelling implies that the quenching fields are same 
with those of resonant tunnelling. 
This exact coincidences are shown in Fig. 4.
The dashed lines denote the EBK energies in the meta-stable well, $E_j^{meta}$.
Note that the cross points with $E^*_0 (=0)$,
$E^*_1$, and $E^*_2$ (the fields of resonant tunnelling)  are 
located exactly on quenching fields
 given by the condition
$\cos \alpha (h) =0$, which is the mechanism for the disappearance of the level-splitting
oscillation.

In conclusion we examine the energy spectrum of the quantum tunnelling
Wess-Zumino spin system by means of the complex periodic orbit theory.
All important features in the spectrum can be understood within this semiclassical
scheme.  Especially, we find out that the absence of the oscillation of the level
splittings at higher magnetic field has its physical origin in the exact cancellation
of quenching with the resonant tunnelling phenomena.
It is more reasonable to conclude that this cancellation is an inherent property
of Wess-Zumino tunnelling system.

We thank Stephen Creagh and C.-H. Kim for useful discussions.

 \newpage
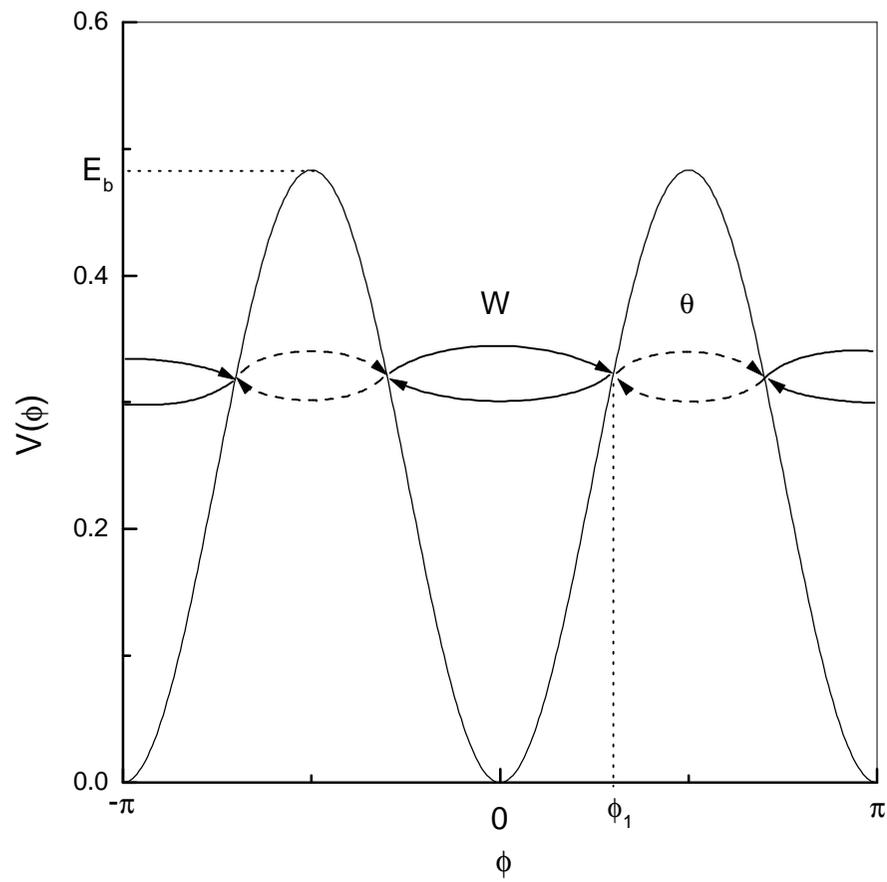
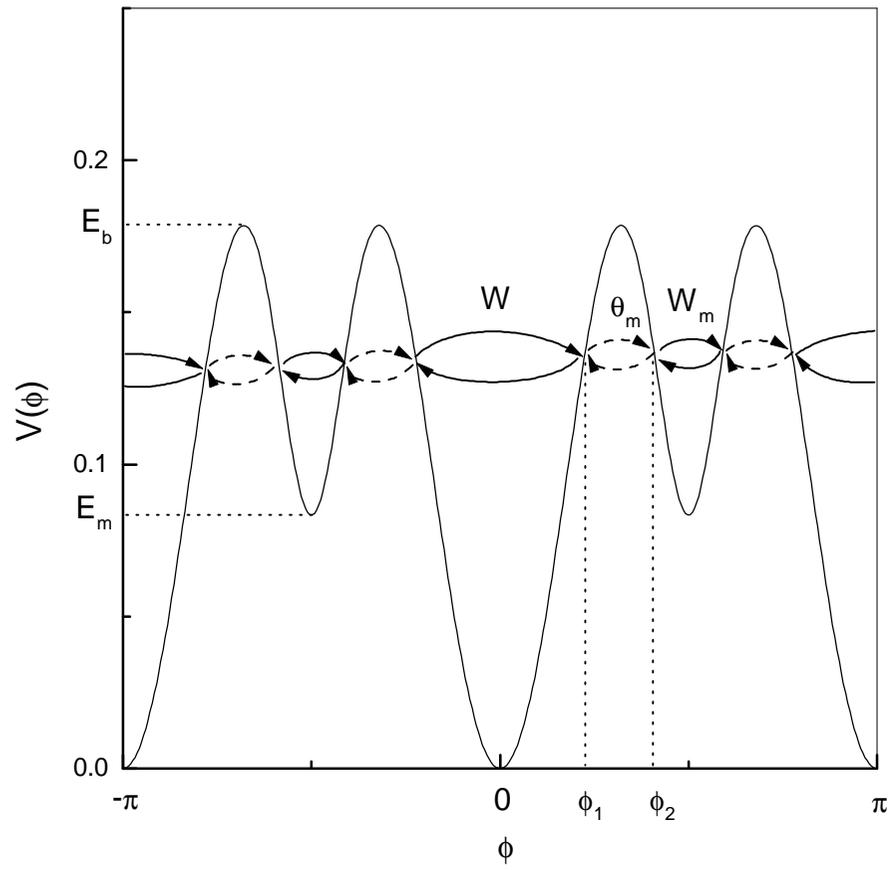
 \begin{figure}
 \caption{The effective potential Eq. (\protect{\ref{pot}}). 
(a) $h < h_1$ case ($\lambda = 0.7$ and $h=0.1$).
The solid and dashed arrows denote the real segments with action $W$
and the tunnelling segments with Euclidean action $\theta$, respectively.
(b) $h > h_1$ case ( $\lambda = 0.7$ and $h=0.5$).
The solid arrows in the meta-stable well represent the path with action $W_m$,
while the dashed arrows do the tunnelling subsegments with Euclidean action
$\theta_m$.
 }
 \label{figure1}
 \end{figure}

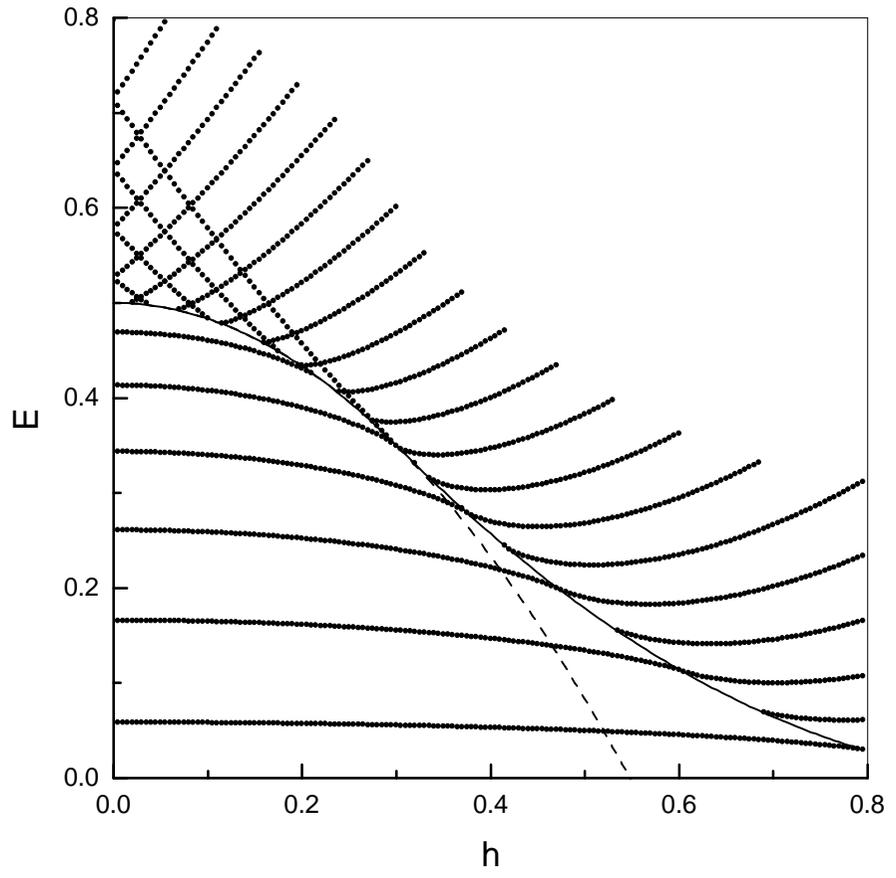
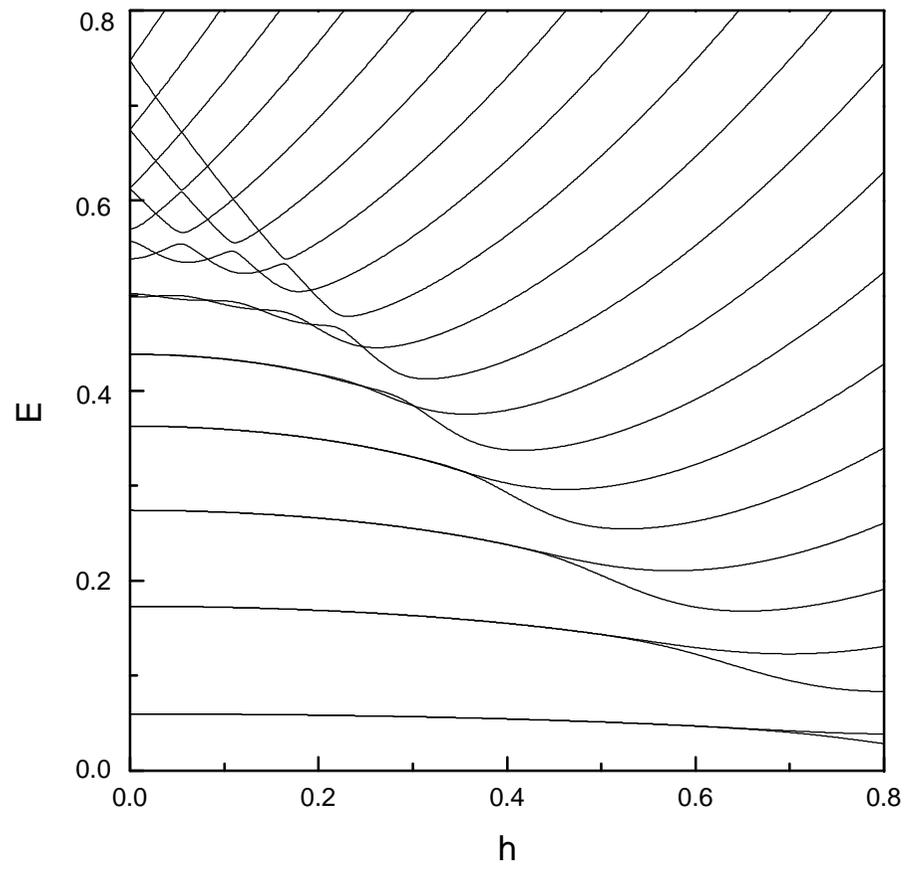
\begin{figure}
\caption{The whole energy spectra for the Hamiltonian Eq. (\protect{\ref{ham}})
with spin value $S=10$ and $\lambda = 0.7$.
(a) The result of the present semiclassical theory. The solid and dashed lines
represent the energy of barrier top $E_b$ and the energy of the local minimum $E_m$,
respectively.
For the energy range of $E < E_b$ we draw only the EBK energies.
(b) The result of direct diagonalization for the Hamiltonian Eq. (\protect{\ref{ham}}).
}
\label{figure2}
\end{figure}

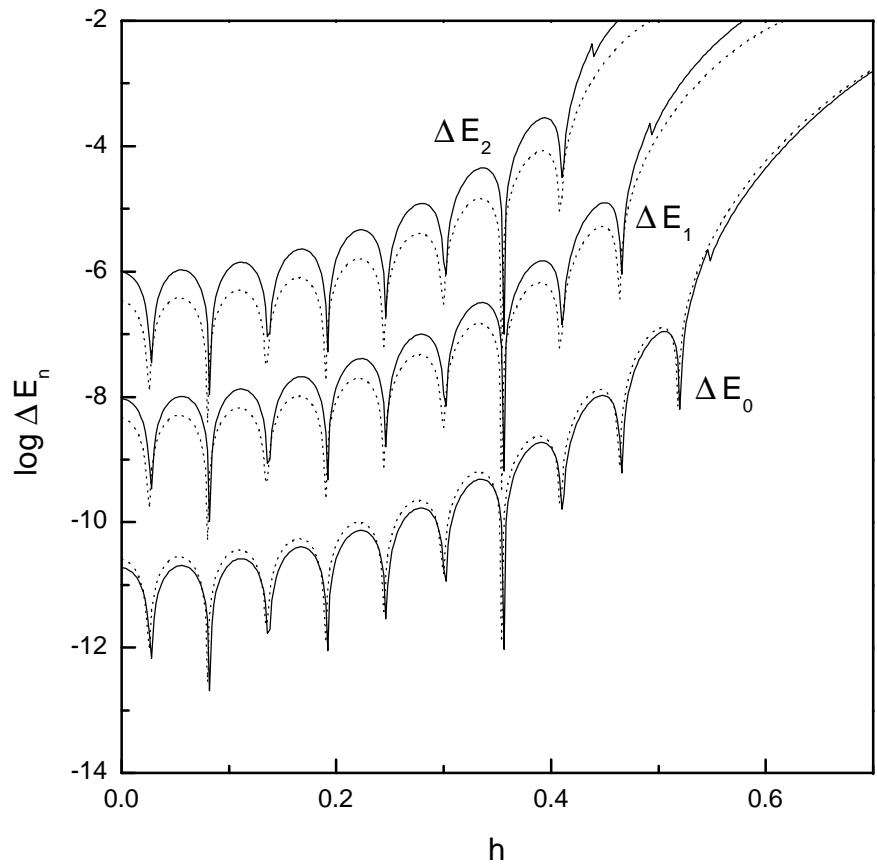
\begin{figure}
\caption{ The energy level splittings $\Delta E_0$, $\Delta E_1$, 
and $\Delta E_2$ in the case of $S=10$ and $\lambda = 0.7$.
The solid lines denote the result of the complex periodic orbit
theory ($\Delta E^s_i$) and the dashed lines do the result of
direct diagonalization ($\Delta E^q_i$).
}
\label{figure3}
\end{figure}

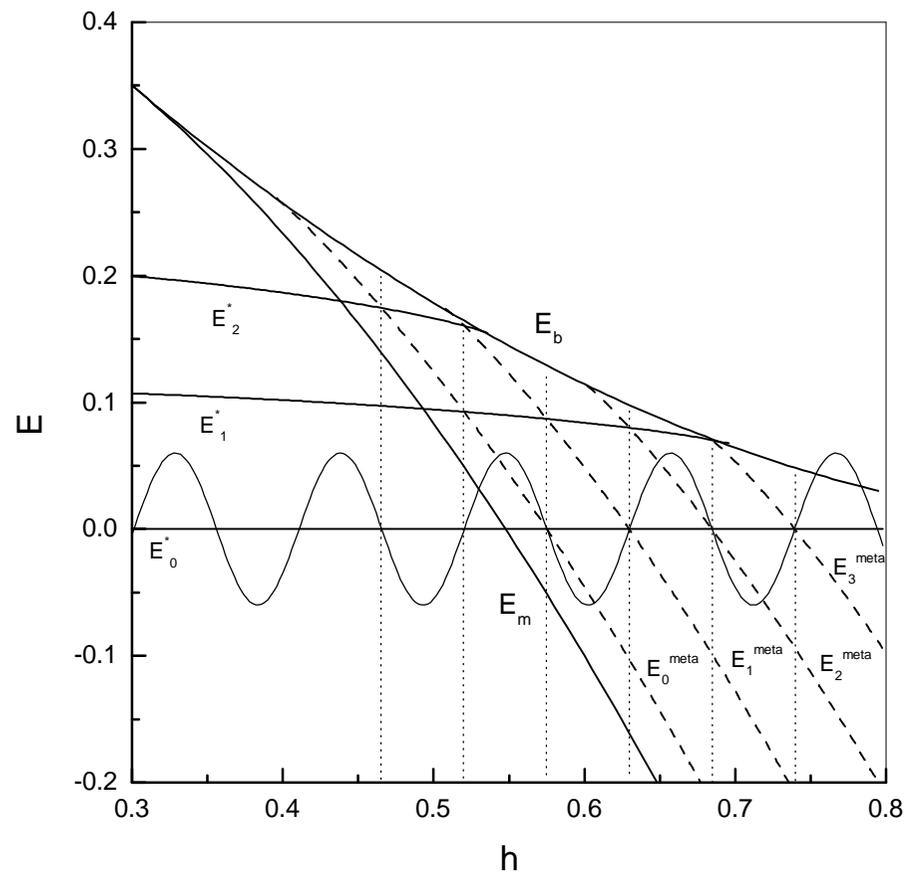
\begin{figure}
\caption{ The EBK energies in the meta-stable well $E_j^{meta}$(dashed lines) and
$E^*_k$(solid lines). The cross points imply the resonant tunnelling.
 The cosine function is $0.06 \cos \alpha (h)$.
 Note that the quenching fields given by $\cos \alpha (h) =0$ are located on the
 same $h$ with the fields of resonant tunnelling.
}
\label{figure4}
\end{figure}

\widetext
\begin{table}
\caption{ Analytic expressions for various actions. $K ( k)$ and $\Pi (\alpha^2, k)$ are 
the complete elliptic integrals of first and third kind, respectively. } 
\begin{tabular}{l}
$W(E)= \frac{2}{\sqrt{A-\lambda B_-}}          
           [ A ( B_- -\frac{1}{\lambda}) \Pi (\alpha_1^2, k_1)   
          + (1-B_-)\Pi(\alpha_2^2,k_1) + \frac{A-\lambda}{\lambda} K (k_1 )  ]$   \\
$\theta(E) =\frac{2}{\sqrt{A-\lambda B_-}}          
           [ \frac{(A-\lambda) ( 1-\lambda B_-)}{\lambda (1-\lambda)} \Pi (\alpha_3^2, k_2) 
          + B_-\Pi(\alpha_4^2,k_2) - \frac{A}{\lambda} K (k_2 )  ]$ \\  
$W_m(E) = \frac{2}{\sqrt{\lambda (1-B_-)B_+}}          
           [ \frac{2 \lambda E - h^2}{\lambda } K ( k_3)    
          -  \Pi(\alpha_5^2,k_3)  
         + \frac{h^2}{\lambda (1-\lambda )} [ \frac{\alpha_5^2}{\alpha_6^2}
         K (k_3) + (1-\frac{\alpha_5^2}{\alpha_6^2} ) \Pi (\alpha_6^2, k_3)  ]]$  \\
$\theta_m(E) =\frac{1}{\sqrt{\lambda (1-B_-)B_+}}          
           [ -\frac{2\lambda E- h^2}{\lambda} K ( k_4)   
          -(1- B_+)\Pi(\alpha_7^2,k_4) + K (k_4 )    
            -\frac{h^2}{\lambda (1-\lambda B_-)}  [ \frac{\alpha_7^2}{\alpha_8^2}
         K (k_4) + (1-\frac{\alpha_7^2}{\alpha_8^2} ) \Pi (\alpha_8^2, k_4)  ]]$  \\
$S_{rot}(E) = \frac{4A}{\lambda \sqrt{(A-\lambda ) B_-}} 
[   \Pi (\alpha_9^2 ,k_5)
- (1-\lambda B_-) \Pi (\alpha_{10}^2,k_5)  ]$    (in the case of real $B_-$) \\
$S_{rot} (E) = \frac{4}{\sqrt{\lambda \mu \nu}} [ \frac{\nu}{\mu -\nu} 
+2E - \frac{h^2}{\lambda} +\frac{h^2(\mu -\nu}{\lambda ( \mu -(1-\lambda)\nu)}] K(k_6) 
-\frac{1}{\sqrt{\lambda \mu \nu}}[ \frac{\mu + \nu}{\mu -\nu}
\Pi (\alpha_{11}^2, k_6)   
+ \frac{h^2 (\mu + (1-\lambda)\nu)}{(1-\lambda)(\mu - (1-\lambda)\nu)}
\Pi (\alpha_{12}^2,k_6) ]$   ( complex $B_-$)\\ \tableline
$ A \equiv \frac{2E}{B_-}$,
 $ B_{\pm} \equiv 
\frac{ (1-h^2 + 2\lambda E) \pm \sqrt{ (1-h^2 + 2\lambda E)^2 - 8 \lambda E } }{2 \lambda}$,  
$k_1^2  \equiv \frac{ B_-(A - \lambda)}{A - \lambda B_-},
\alpha_1^2 \equiv  \frac{\lambda B_- (A - 1)}{A - \lambda B_-}, 
\alpha_2^2  \equiv  B_-$  \\
$k_2^2    \equiv  \frac{A(1-B_-)}{A-\lambda B_-},
\alpha_3^2 \equiv  \frac{\lambda (1- A) ( 1-B_-)}{(1-\lambda )(A - \lambda B_-)},
\alpha_4^2  \equiv  1-B_-$,  
$k_3^2    \equiv  \frac{B_-(1-B_+)}{B_+(1- B_-)}, 
\alpha_5^2 \equiv  -\frac{ 1-B_+}{B_+},  
\alpha_6^2  \equiv  -\frac{(1-B_+)}{B_+(1-\lambda)}$ \\
$k_4^2    \equiv  \frac{(B_+-B_-)}{B_+(1- B_-)}, 
\alpha_7^2 \equiv  \frac{ B_+-B_-}{1-B_-},  
\alpha_8^2  \equiv  \frac{(B_+-B_-)}{B_+(1-\lambda B_-)}$, 
$k_5^2 \equiv \frac{A- \lambda B_-}{ (A - \lambda ) B_- }$,
$\alpha_9^2 \equiv -\frac{\lambda }{A-\lambda}$,
$\alpha_{10}^2 \equiv \frac{\lambda ( A-1)}{A-\lambda}$ \\
$\mu^2 = (1- {\rm Re}B_-)^2+({\rm Im}B_-)^2, 
\nu^2 = ({\rm Re} B_-)^2 + ({\rm Im}B_-)^2$, 
$k_6^2 \equiv \frac{1-(\mu - \nu)^2}{4\mu \nu}$, 
$\alpha_{11}^2 \equiv -\frac{(\mu-\nu)^2}{4\mu \nu}$,
$\alpha_{12}^2 \equiv -\frac{(\mu-(1-\lambda)\nu )^2}{4(1-\lambda)\mu \nu}$
\\
\end{tabular}
\end{table}

\end{document}